\documentclass[preprint,5p,number,sort&compress]{elsarticle}
\usepackage{dcolumn}
\usepackage{bm}
\usepackage{upgreek}
\usepackage{graphicx}
\usepackage{hyperref}
\usepackage[load-configurations=abbreviations,detect-all=true]{siunitx}
\usepackage[version=3]{mhchem}
\usepackage[mathlines]{lineno}
\usepackage[usenames,dvipsnames,svgnames,table]{xcolor} 
\usepackage[normalem]{ulem} 
\usepackage{transparent}

\usepackage{xspace}

\newcommand{\TM}{\emph{Topmetal}\xspace}

\newcommand{\TMIIm}{\mbox{\emph{Topmetal-II\raise0.5ex\hbox{-}}}\xspace}

\newcommand{\Vtp}{\ensuremath{V_\text{TP}}\xspace}
\newcommand{\Vs}{\ensuremath{V_\text{step}}\xspace}

\newcommand{\Cinj}{\ensuremath{C_\text{inj}}\xspace}

\newcommand{\CinjVal}{\SI{5.5}{fF}\xspace}

\newcommand{\Vth}{\ensuremath{V_{\text{th}}}\xspace}
\newcommand{\Vthg}{\ensuremath{V_{\text{th}g}}\xspace}
\newcommand{\Vthp}{\ensuremath{V_{\text{th}p}}\xspace}

\DeclareSIUnit\keV{keV}
\DeclareSIQualifier\ee{ee}
\DeclareSIQualifier\nr{nr}
\DeclareSIUnit\pe{p.e.}

\DeclareMathOperator*{\erf}{erf}

\newcommand{\sym}[1]{\texttt{#1}}

\journal{Nuclear Instruments and Methods in Physics Research A}

\begin{document}

\begin{frontmatter}

\title{Detailed study of the column-based priority logic readout of \TMIIm CMOS pixel direct charge sensor}

\author[ccnu]{Mangmang An}
\author[ccnu]{Chufeng Chen}
\author[ccnu]{Chaosong Gao}
\author[lbnl]{Mikyung Han}
\author[ccnu]{Guangming Huang}
\author[ccnu]{Rong Ji}
\author[ccnu]{Xiaoting Li}
\author[lbnl]{Yuan Mei\corref{cor0}}\ead{ymei@lbl.gov}
\author[ccnu]{Hua Pei}
\author[ioa]{Quan Sun}
\author[ccnu]{Xiangming Sun\corref{cor0}}\ead{xmsun@phy.ccnu.edu.cn}
\author[ccnu]{Kai Wang}
\author[ccnu]{Le Xiao}
\author[ccnu]{Ping Yang}
\author[ccnu]{Wei Zhang}
\author[ccnu]{Wei Zhou}

\address[ccnu]{PLAC, Key Laboratory of Quark and Lepton Physics (MOE), Central China Normal University, Wuhan, Hubei 430079, China}
\address[lbnl]{Nuclear Science Division, Lawrence Berkeley National Laboratory, Berkeley,
  California 94720, USA}
\address[ioa]{Institute of Acoustics, Chinese Academy of Sciences, Beijing 100190, China}

\cortext[cor0]{Corresponding author}

\begin{abstract}

  We present the detailed study of the digital readout of \TMIIm CMOS pixel direct charge sensor.
  \TMIIm is an integrated sensor with an array of $72\times72$ pixels each capable of directly
  collecting external charge through exposed metal electrodes in the topmost metal layer.  In
  addition to the time-shared multiplexing readout of the analog output from Charge Sensitive
  Amplifiers in each pixel, hits are also generated through comparators with individually DAC
  settable thresholds in each pixel.  The hits are read out via a column-based priority logic
  structure, retaining both hit location and time information.  The in-array column-based
  priority logic is fully combinational hence there is no clock distributed in the pixel array.
  Sequential logic and clock are placed on the peripheral of the array.  We studied the detailed
  working behavior and performance of this readout, and demonstrated its potential in imaging
  applications.

\end{abstract}

\end{frontmatter}

\section{Introduction}\label{sec:intro}

Highly pixelated sensors such as CMOS image sensors and Monolithic Active Pixels Sensors (MAPS),
have been successfully deployed in various fields.  In many nuclear/particle physics
applications, the traditional ``rolling shutter'' style readout (time-shared multiplexing) is
orders of magnitude slower than what is required for signal/data acquisition in order to achieve
the physics goals.  Therefore, many novel readout schemes, designed to access the information
collected in pixels in the sensor faster, have been developed over the years.  These schemes
often exploit the characteristics of signal distribution among pixels, such as the sparseness or
clustering in space and time of pixel hits, and the similarity in amplitudes.  Notable examples
include column-based readout\cite{Millaud1995, Yang201561, GarciaSciveres2011S155} and row-based
compression \cite{HuGuo2009, HuGuo2010480}.

We implemented a column-based priority logic readout in a prototype pixel sensor called
\TMIIm\cite{An2016}, aimed at improving the latency between a pixel hit and the availability of
data off the chip.  \TMIIm is implemented in a \SI{0.35}{\micro m} CMOS process.  It features a
$72\times72$ pixel array with \SI{83.2}{\micro m} pixel pitch.  Pixels in \TMIIm are sensitive to
external charges arrived at an exposed metal electrode in the topmost layer in each pixel.  Pixel
hits are generated by pixel-local comparators with tunable thresholds.  We designed \TMIIm
towards achieving both low analog noise and low latency in digital readout.  The analog front-end
achieved a $<\SI{14}{e^-}$ Equivalent Noise Charge (ENC)\cite{An2016} per pixel.  For the digital
circuitry, we chose a scheme that is clock-less (fully combinational) in the pixel array to
minimize the potential interference from digital activities (flips).  Digital activities only
happen when the sensor receives hits.  The in-array column-based combinational logic drives the
address of the pixel that is hit to the edge of the array immediately upon a hit, which minimizes
the latency.  A sequential logic (with clock) is employed to sense the hit location and time at
the edge of the array then ship such information off the sensor.

A detailed study of the analog characteristics of \TMIIm is reported in \cite{An2016}.  This
paper focuses on the details of operation and performance of the digital readout.

\section{Sensor structure and operation}\label{sec:so}

\begin{figure*}[!htb]
  \centering
  \begin{minipage}[c]{0.4\linewidth}
    \vspace{7ex}
    \includegraphics[width=\linewidth]{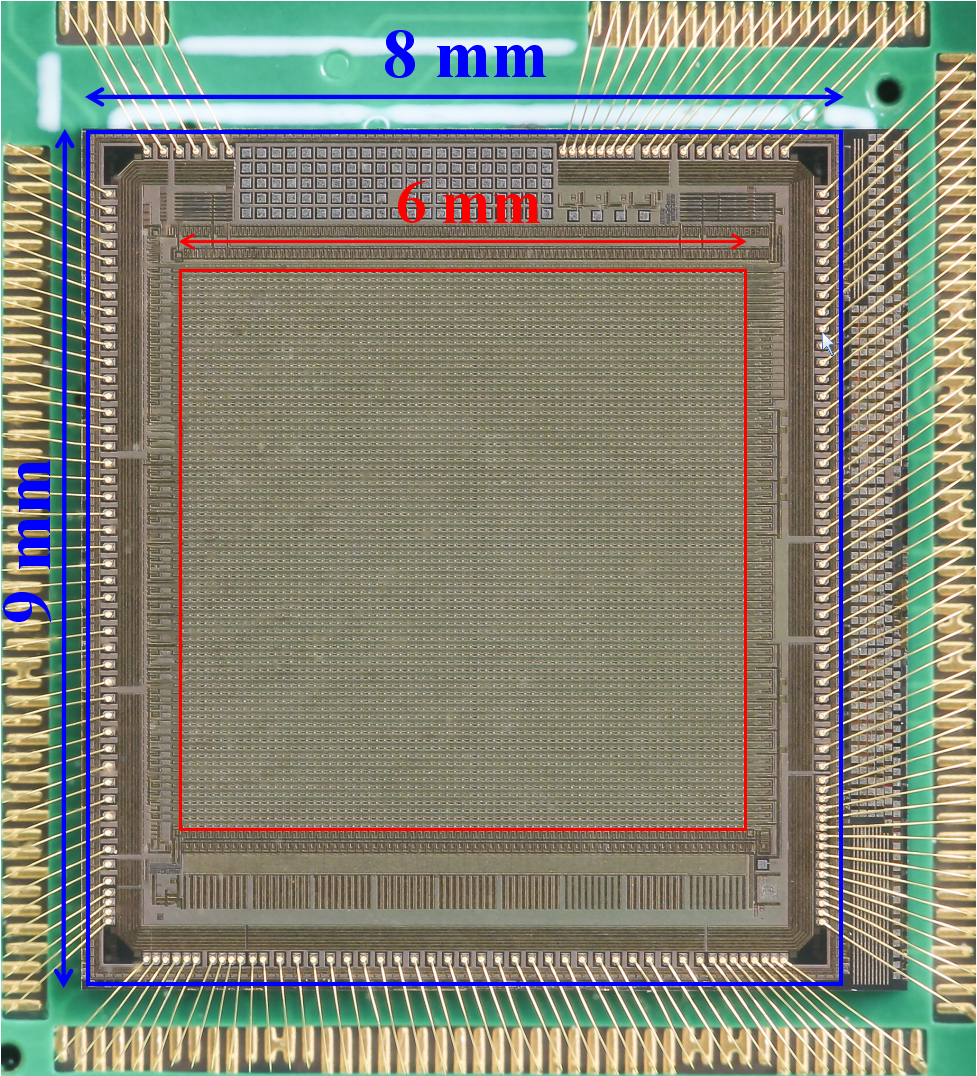}
  \end{minipage}%
  \begin{minipage}[c]{0.6\linewidth}
    \vspace{5ex}
    \includegraphics[width=\linewidth]{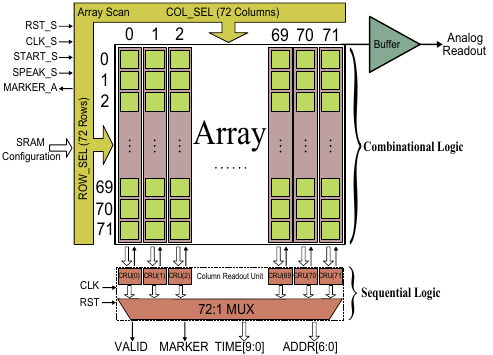}
  \end{minipage}
  \caption{Photograph of a \TMIIm sensor (left) and its top-level block diagram (right).  The
    chip is $8\times\SI{9}{mm^2}$ (blue box) in size, in which a $6\times\SI{6}{mm^2}$ charge
    sensitive area (red box) containing $72\times72$ pixels is located in the center of the
    sensor.  Major functional units are shown in the top-level block diagram.  The digital logic
    inside of the array is entirely combinational without a clock.  Sequential logic driven by a
    clock is placed at the edge of the array.}
  \label{fig:TMIImOverall}
\end{figure*}

A \TMIIm sensor, as shown in Fig.~\ref{fig:TMIImOverall}, contains an array of $72\times72$
sensitive pixels occupying a $6\times\SI{6}{mm^2}$ area.  Each pixel has an exposed metal patch
in the topmost layer that can directly collect charge.  The charge signal is amplified by a
Charge Sensitive Amplifier (CSA) in each pixel (Fig.~\ref{fig:pixStr}).  Light illumination also
results in charge signal, which is then amplified by the same CSA.  The amplified charge signal
is accessible through two channels.  The analog voltage signal is read out through a ``rolling
shutter'' style time-shared multiplexer controlled by the array scan unit.  A digital hit signal
is generated by an in-pixel comparator with per-pixel adjustable threshold, then read out through
the column-based priority logic.

\subsection{Column-based priority logic readout}

The overall readout has two parts: an in-array combinational logic (Fig.~\ref{fig:pixStr}, purple
dashed box) and a sequential logic (Fig.~\ref{fig:pixStr}, yellow dashed box) at the bottom edge
of the array.  The combinational logic consists of a Priority Logic (PL) in each pixel and an
Address Bus (AB) in each column.  The sequential logic includes a Column Readout Unit (CRU)
placed at the bottom edge of each column and a multiplexer (MUX) congregating the outputs of all
CRUs.  CRUs monitor the address changes on the ABs.  CRUs and the MUX are synchronous to shared
clock \sym{CLK} and reset \sym{RST} signals.

\subsubsection{Pixel hits and Priority Logic (PL)}

\begin{figure*}[!htb]
  \centering
  \includegraphics[width=\linewidth]{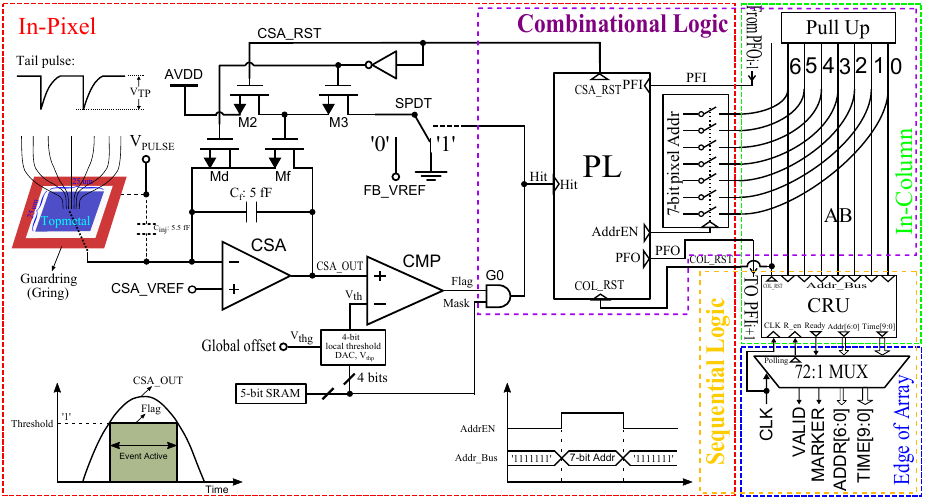}
  \caption{Digital readout pathway from a single pixel to the edge of array.  Structures at
    pixel, column and edge-of-array levels are indicated in the red, green and blue dashed boxes,
    respectively.  A hit generated by \sym{CSA\_OUT} rising above the threshold (left-side inset
    plot) propagates through the in-pixel priority logic that drives the column address bus
    (\sym{Addr\_Bus}) by asserting \sym{AddrEN} (right-side inset plot), which in turn is read
    out by the CRU and the MUX.  The in-pixel logic and the column Address Bus (AB) are entirely
    combinational (purple dashed box).  The CRU \& MUX shown in the yellow box are sequential
    logic driven by a common clock.}
  \label{fig:pixStr}
\end{figure*}

A schematic view of the circuit in a single pixel is shown in the red dashed box in
Fig.~\ref{fig:pixStr}.  The exposed \TM electrode is directly connected to the input of the CSA.
A ring electrode (\sym{Gring}), which is in the same topmost metal layer as the \TM, surrounds
the \TM while being isolated from it.  The stray capacitance between the \sym{Gring} and the \TM,
$\Cinj\approx\CinjVal$, is a natural test capacitor that allows applied pulses on \sym{Gring} to
inject charge into the CSA.  The CSA with $C_f\approx\SI{5}{fF}$ converts the injected charge to
voltage signal (\sym{CSA\_OUT}) and feeds it into the comparator.  The comparator compares
\sym{CSA\_OUT} to a threshold (\Vth) set by a pixel-local 4-bit DAC (\Vthp) on top of a common
offset \Vthg that is globally adjustable.  $\Vth{_i}=\Vthp{_i}+\Vthg$, where $i$ is the index of
pixel in the array.  The step size of all the 4-bit DACs is globally adjustable as well.  The
pixel-local 4-bit DAC is intended for compensating the threshold dispersion of the comparator
across the entire array.  The CSA and the comparator are constantly active.

Upon an event that \sym{CSA\_OUT} surpasses the threshold \Vth, the comparator asserts
$\sym{Flag}=1$, which propagates to an AND gate \sym{G0} (Fig.~\ref{fig:pixStr}).  The other
input of \sym{G0}, \sym{Mask}, is used for disabling pixels from responding to hits digitally.
This feature is exploited during the digital readout tests and imaging demonstration.  The
\sym{Mask} together with the 4-bits for DAC in each pixel are set by a pixel-local 5-bit SRAM.
Writes to SRAMs are synchronous to array scan.  SRAMs were chosen over Flip-Flops to save floor
space.

\begin{figure}[!htb]
  \centering
  \flushleft{(a)}\\
  \includegraphics[width=\linewidth]{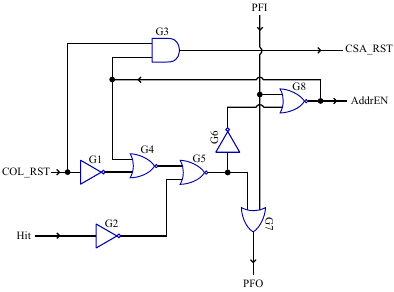}\\
  \vspace{-4ex}
  \flushleft{(b)}\\
  \begin{tabular}{c c c | c | c | c c}
    \hline\hline
    \multicolumn{3}{c|}{Input} & \multicolumn{2}{c|}{\sym{AddrEN}} & \multicolumn{2}{c}{Output}\\
    \hline
    \sym{PFI} & \sym{Hit} & \sym{COL\_RST} & Initial & Final & \sym{PFO} & \sym{CSA\_RST}\\
    \hline
    0 & 1 & 1 & 0 &                   0 & 0 & 0      \\
    0 & 1 & 1 & 1 &                   1 & 1 & 1      \\
    0 & 1 & 0 & X &                   1 & 1 & 0      \\
    0 & 0 & X & X &                   0 & 0 & 0      \\
    1 & X & X & X &                   0 & 1 & 0      \\
    \hline
  \end{tabular}
  X$=$do not care.
  \caption{(a) Schematic of in-pixel Priority Logic (PL) circuitry.  (b) Truth table of PL.}
  \label{fig:PL}
\end{figure}

When \sym{G0} outputs 1, a hit is generated ($\sym{Hit}=1$) and the PL module is notified.  Each
PL is a fully combinational logic that controls the reset (\sym{CSA\_RST}) of the CSA upon the
readout of a hit and drives the hit information through the column structure.  The internal
structure of PL and its truth table are shown in Fig.~\ref{fig:PL}.

\subsubsection{Column-wise priority chain and Address Bus (AB)}

The priority logic signals propagate in columns.  For the $i$th pixel, its \sym{PFI}$_i$ is
connected to the previous ($(i-1)$th) pixel's \sym{PFO}$_{i-1}$, and its \sym{PFO}$_i$ is fed
into the next ($(i+1)$th) pixel's \sym{PFI}$_{i+1}$.  Pixels in the same column are daisy-chained
in this fashion.  Every pixel in the same column has a unique hard-coded 7-bit address in the
form of pull-down switches.  Encoded pull-down switches are connected to the column-shared
Address Bus (AB) (green dashed box in Fig.~\ref{fig:pixStr}).  AB is weakly pulled up to all high
by default.  When \sym{AddrEN} becomes active in a pixel, said pixel pulls down the AB to its own
unique address.  The topmost pixel (0th) in a column has \sym{PFI}$_0=0$.  If there is no hit in
any pixel (\sym{Hit} $=0$), the \sym{PFO} output is forced to 0 by \sym{G2}, \sym{G5} \&
\sym{G7}, which dictates that every pixel in the column has $\sym{PFI}=\sym{PFO}=0$.  When there
is no active \sym{COL\_RST} sent from the CRU module to every pixel in the column simultaneously,
the outputs of \sym{G3} \& \sym{G4} are forced to 0.  Once a pixel (e.g.\ $i$th) gets a hit, due
to the effects of \sym{G2}, \sym{G5} \& \sym{G7}, $\sym{PFO}_i=1$.  Forced by \sym{G7}, all
pixels below the $i$th pixel (denoted by $j$th, $j>i$) will have $\sym{PFI}_j=\sym{PFO}_j=1$.
Forced by \sym{G8}, any pixel with $\sym{PFI}=1$ won't enable \sym{AddrEN} even if it gets a hit.
The above described logic forms a column-wise priority chain: only the pixel with a hit that has
the lowest $i$ (highest priority) enables its \sym{AddrEN}, and it disables all the pixels lower
in the chain from asserting their individual addresses on the AB.  Therefore, AB is pulled down
by only one pixel (the highest priority pixel with a hit) at any given time so that no race
condition rises on the AB.

\subsubsection{Column Readout Unit (CRU)}

Each priority chain (column) is terminated by a Column Readout Unit (CRU) at the bottom of the
column.  CRU monitors the Address Bus (AB) and validates the address change, then records the
7-bit address \& 10-bit time stamp for the corresponding hit pixel.  Upon the read of a hit, the
CRU asserts $\sym{COL\_RST}=1$, which is fed back simultaneously to all the pixels in the column.
Only the pixel that is pulling on the bus will respond to \sym{COL\_RST} (see Fig.~\ref{fig:PL}),
which results in the analog reset of the CSA ($\sym{CSA\_RST}=1$), the removal of hit, and the
release of the bus.  When the bus is successfully released, the address seen by the CRU returns
to all high.  The CRU senses such condition and outputs $\sym{Ready}=1$.  It indicates that a hit
has been registered in the column and has not yet been read by the MUX.  \sym{COL\_RST} and
\sym{Ready} are kept high until this CRU is read by the MUX.  \sym{R\_en} is set to high by the
MUX when it reads the associated CRU.

\subsubsection{Multiplexer (MUX)}

As shown in the blue dashed box of Fig.~\ref{fig:pixStr}, a digital multiplexer (MUX) polls the
status of each CRU sequentially, advancing at the falling edge of each clock cycle.  It picks up
the valid addresses and time stamps for the hit pixels from each CRU, then ships them off the
sensor.  A \sym{MARKER} signal is asserted when the 0th column is polled to indicate the start of
a frame.  The index of the column being read can be calculated externally referencing to
\sym{MARKER}.  A \sym{VALID} signal is asserted when the column being read has a hit.  The
address and time stamp outputs are valid only when $\sym{VALID}=1$.

\subsection{Readout operation and timing}

\begin{figure}[!htb]
  \centering
  \includegraphics[width=\linewidth]{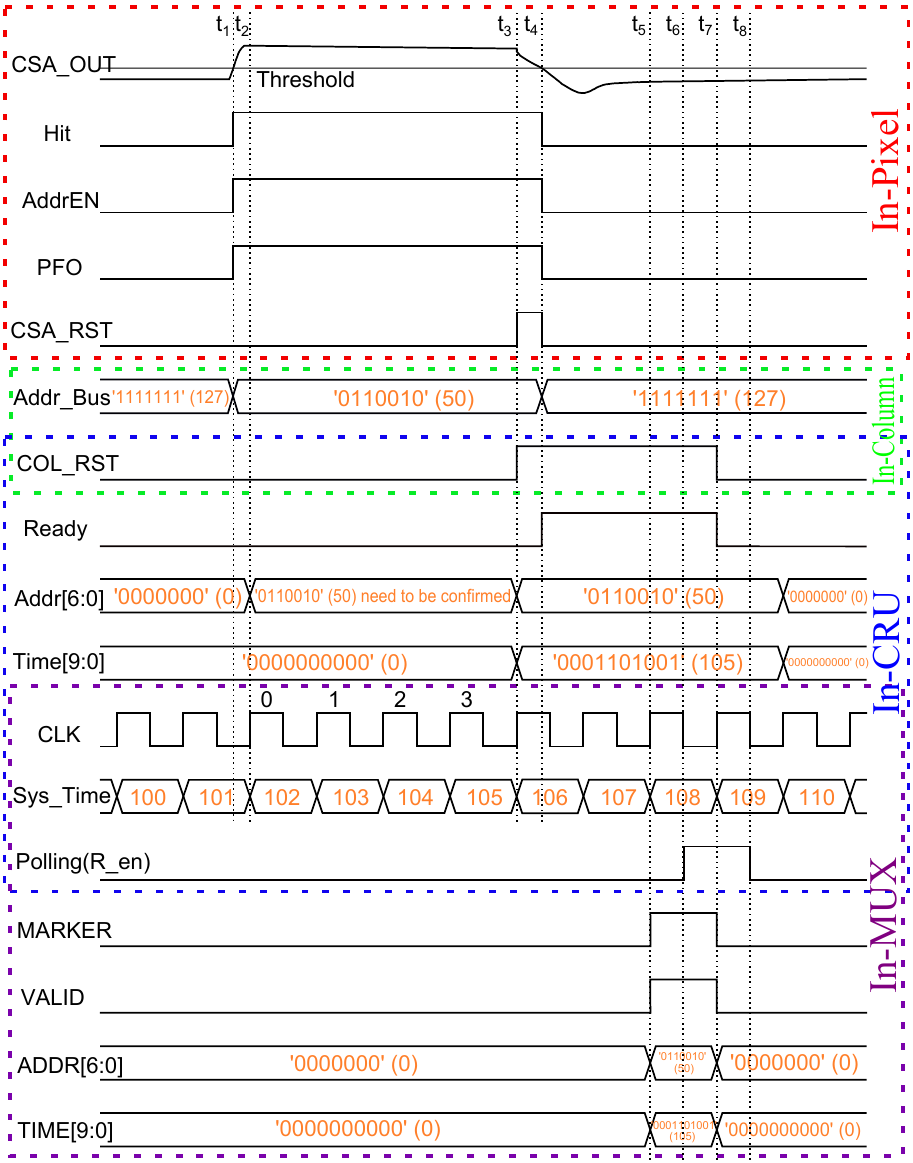}
  \caption{Timing diagram of relevant signal activities during a hit and its readout.  In-pixel,
    in-column, in-CRU and in-MUX signals are indicated in red, green, blue and purple dashed
    boxes, respectively.}
  \label{fig:PriLogicSeq}
\end{figure}

A timing diagram of the readout process of a valid hit is shown in Fig.~\ref{fig:PriLogicSeq}.
It is assumed that there is only one pixel at Row 50, Column 0 is hit and the system counter has
an initial value of $\sym{Sys\_Time[9:0]}=0001100100_2 (100_{10})$.  We also set \sym{Mask}$=1$
to enable the pixel response to hits.

Charges arrive at $t_1$, causing the CSA output to exceed the threshold of the comparator,
resulting in $\sym{Flag}=1$.  Since $\sym{Mask}=1$, a hit is generated ($\sym{Hit}=1$); hence,
the single-pole-double-throw (\sym{SPDT}) switch (Fig.~\ref{fig:pixStr}) grounds the gate of Mf
from its original bias \sym{FB\_VREF} so the CSA maximally retains the charge signal.  As
$\sym{PFI}=0$, following the logic in Fig.~\ref{fig:PL}, \sym{PFO} and \sym{AddrEN} become 1
accordingly.  At this moment ($t_1$), the Address Bus (AB) is pulled to the address of this pixel
as well ($\sym{Addr\_Bus}=0110010_2 (50_{10})$).  At $t_2$ (rising edge of the clock in the CRU),
the CRU senses the address change and outputs $\sym{Addr[6:0]}=0110010_2 (50_{10})$, and waits
for 4 clock cycles to confirm that the address change is not a transient phenomenon.  At the end
of the waiting period, $t_3$, the CRU latches the address value and the time stamp from the
system counter $\sym{Time[9:0]}=\sym{Sys\_Time[9:0]}=0001101001_2 (105_{10})$.  It also sends a
reset signal \sym{COL\_RST}$=1$ back to the column.  Although \sym{COL\_RST} is sent to every
pixel in the column, forced by \sym{G3} in Fig.~\ref{fig:PL} (a), only the pixel that is pulling
the AB and is being read out will respond to the reset.  The reset sets $\sym{CSA\_RST}=1$, which
turns on the feedback transistor Mf, discharging $C_f$ so that the CSA output comes down towards
the baseline.  At $t_4$, the CSA output falls below the threshold, causing $\sym{Hit}=0$ hence
$\sym{AddrEN}=0$ and $\sym{PFO}=0$.  Once $\sym{AddrEN}=0$, CSA reset is done
($\sym{CSA\_RST}=0$) and \sym{Addr\_Bus} returns to all high.  The CRU also sets $\sym{Ready}=1$
indicating there is a valid hit waiting to be read.  Both the \sym{COL\_RST} and \sym{Ready} are
removed when the CRU is polled at $t_7$.  The time between $t_4$ and $t_7$ is non-deterministic
and can be as high as 72 clock cycles.  During $t_5\sim t_7$, the MUX is Polling the CRU and
shipping the data ($\sym{ADDR[6:0]}=0110010_2 (50_{10})$ and
$\sym{TIME[9:0]}=0001100100_2 (105_{10})$) off the sensor.  Since this pixel is in the 0th
column, besides generating a $\sym{VALID}=1$, a synchronous \sym{MARKER} is also simultaneously
asserted.  As the signal \sym{Polling} (\sym{R\_en}) is driven by the falling edge of the clock,
it has a half-clock-cycle delay behind the \sym{MARKER}; therefore, it's high from $t_6$ to
$t_8$.

When multiple pixels in the same column are hit simultaneously, the logic reads out and resets
the hit pixels sequentially following their priorities in descending order.  When a
higher-priority hit pixel is waiting to be polled, the CRU keeps the \sym{COL\_RST} high.  When
$\sym{COL\_RST}=1$, the \sym{G1}\&\sym{G4} ensures that the next-priority hit pixel holds its
$\sym{AddrEN}=\sym{CSA\_RST}=0$ until the \sym{COL\_RST} is removed.  No hit is missed.  However,
due to this behavior, the CRU cannot respond to the next-priority hit in real-time, which causes
the loss of time information for less prior hits.  As shown in Fig.~\ref{fig:TimeInfo}, the time
stamps are only accurate for the pixels with the highest priority.

\section{Measurements and experimental results}\label{sec:mer}

Controlled signal injections, in the form of test pulses applied on the guard ring (\sym{Gring})
and LED pulsed light illumination, were used to measure the thresholds of every pixel and to
demonstrate the imaging capability of the sensor.

\subsection{Threshold and noise}

We applied a repetitive tail pulse with an amplitude \Vtp on \sym{Gring} (see the top-left inset
in Fig.~\ref{fig:pixStr}).  An equivalent negative charge $Q_i=\Cinj\times\Vtp$ is injected at
every falling edge of the pulse into the CSA in every pixel simultaneously.  The response
amplitude of the CSA is expected to be $\Vtp\cdot(\Cinj/C_f)\approx\SI{16.5}{mV}$, subject to a
small variation due to uncertainties in the capacitance.  The CSA responds to both positive and
negative charges equally well; however, only the negative equivalent charge can bring the
\sym{CSA\_OUT} above the threshold to generate hits.  Also, we would like to avoid undershoots of
the CSA output due to positive charge injections; therefore, tail pulses are chosen over a square
wave.  The repetition rate of tail pulses is chosen to be low enough such that all the hit pixels
have sufficient time to be readout and reset before the next pulse arrives.

\begin{figure}[!htb]
  \centering
  \includegraphics[width=\linewidth]{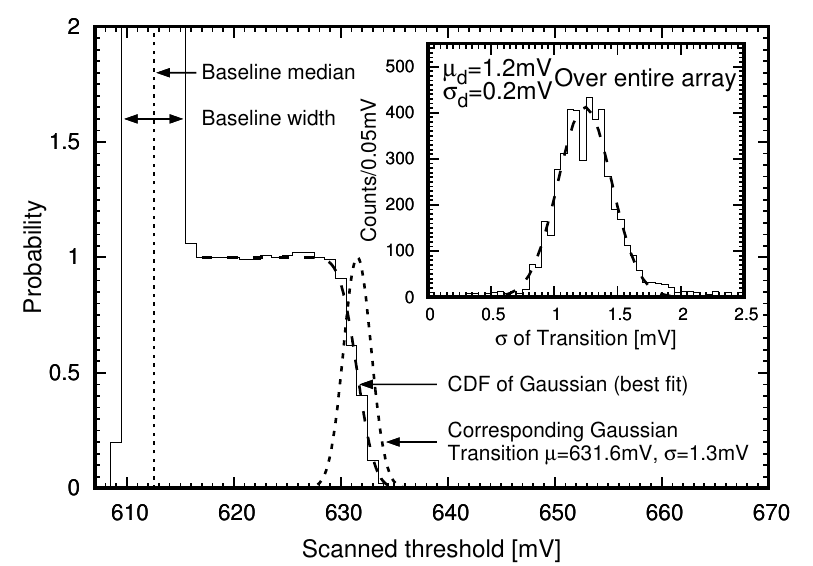}
  \caption{Threshold scan and response parameterization.  A representative S-Curve of one pixel
    is shown.  The baseline median and width are determined with the $\text{probability}>2$ part
    of the curve (see text for details).  The transition region is fitted with the CDF of
    Gaussian to determine its mean ($\mu$) and width ($\sigma$).  Inset shows the distribution of
    the width of the transition ($\sigma$) across all the pixels in the array.}
  \label{fig:S-Curve}
\end{figure}

As shown in Fig.~\ref{fig:S-Curve}, an S-Curve for a single pixel is obtained by scanning the
threshold while recording the corresponding probability for the discriminator and the subsequent
logic to register a hit given a test pulse on the \sym{Gring}.  The threshold is gradually
lowered from well above the signal height where hit $\text{probability}=0$.  When the threshold
is close to the injected signal height, a characteristic tapered transition from probability 0 to
1 due to noise appears.  When the threshold is close to the baseline, the logic registers a hit
every cycle regardless of the injected signal pulse; therefore, the computed probability is
bogusly well above 1.  When the threshold is well below the baseline, the logic saturates and
outputs no hit, although internally the discriminator constantly outputs 1.  We determine the
median and width of the baseline using the $\text{probability}>2$ part of the curve.  We fit the
transition part using the Cumulative Distribution Function (CDF) of Gaussian,
$f(x)=\frac{1}{2}\left[1-\erf\left(\frac{x-\mu}{\sigma\sqrt{2}}\right)\right]$, to determine the
mean ($\mu$) and width ($\sigma$) of the transition.

\begin{figure}[!htb]
  \centering
  \includegraphics[width=\linewidth]{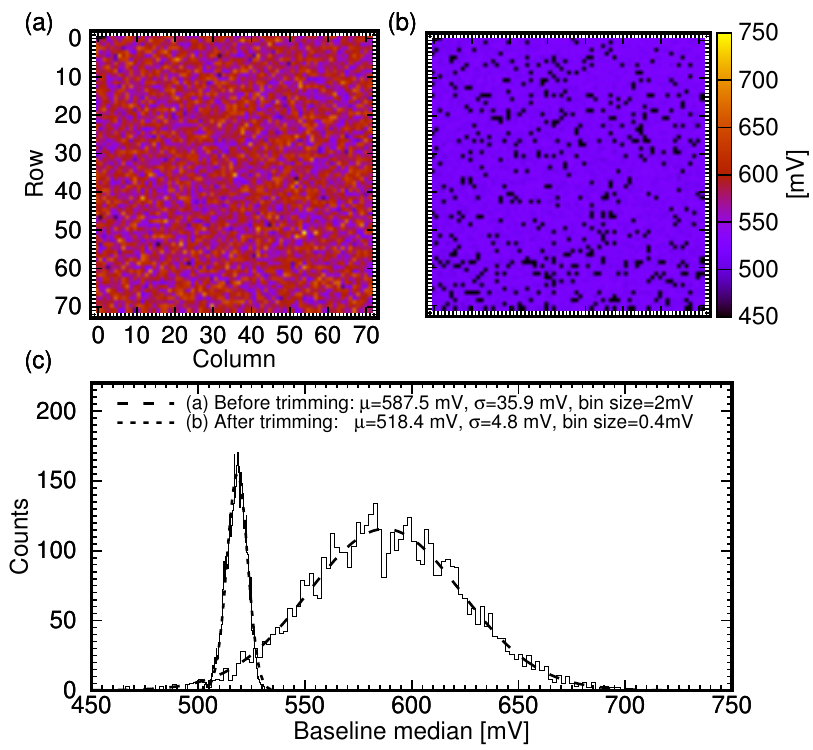}
  \caption{Baseline median distribution across the array in 2D (a, b) and 1D histogram (c) before
    (a and dashed line in c) and after (b and dotted line in c) trimming with a 4-bit DAC in
    every pixel.  The baseline median is defined in Fig.~\ref{fig:baseline_mid}.  The reduction
    of spread ($\sigma$) of the baseline median distribution is clearly visible with the
    trimming.  The shift of the mean is unimportant.  Black points in (b) indicates the pixels
    are disabled.}
  \label{fig:baseline_mid}
\end{figure}

\begin{figure}[!htb]
  \centering
  \includegraphics[width=\linewidth]{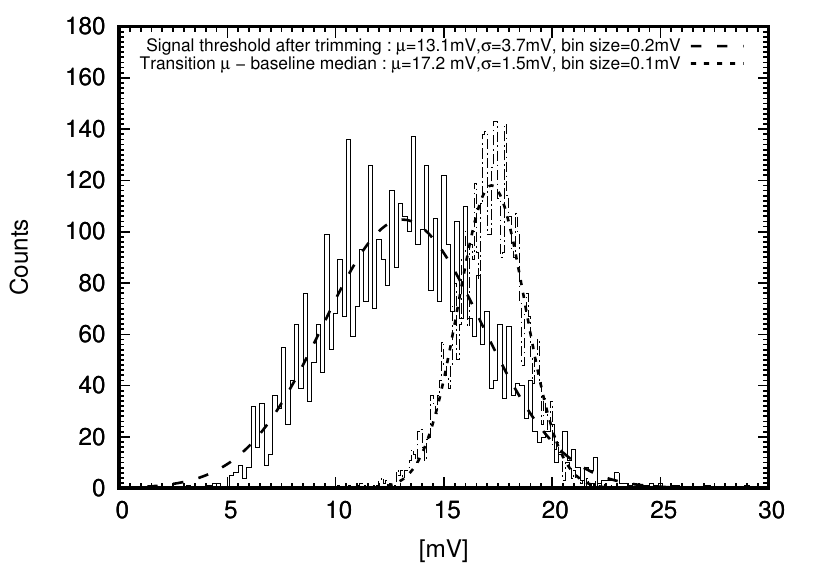}
  \caption{The dotted line shows the distribution of the distance from the transition ($\mu$) to
    the baseline median of an S-Curve (Fig.~\ref{fig:baseline_mid}), which equals to the CSA
    output amplitude response to the injected pulse height.  The dashed line shows the
    distribution of signal threshold in the array with the optimized trimming settings.}
  \label{fig:Threshold}
\end{figure}

The above described procedure is repeated for every pixel in the array.  4-bit DACs are set to 0
for all pixels while the global \Vthg is varied to achieve the threshold scan.  Since the test
pulse on \sym{Gring} injects charges into all pixels in the array simultaneously, to avoid
unnecessary traffic in the priority chain, we used the \sym{Mask} to enable one row at a time, so
that the column readout will read hits from only one pixel in each column.  Through the threshold
scan procedure for the entire array, we extracted the baseline and transition's location and
width from recorded S-Curves of every pixel.  The width ($\sigma$) of transition, which is an
indicator of the noise of CSA output presented to the comparator, has a mean value of
\SI{1.2}{mV} (see the inset in Fig.~\ref{fig:S-Curve}).  It is consistent with the analog noise
measurement reported in \cite{An2016}.  The baseline median distribution of the array is shown in
Fig.~\ref{fig:baseline_mid}(a) and (c).  The distribution of transition $\mu-$baseline median is
shown in Fig.~\ref{fig:Threshold}.  Although the baseline median has a large dispersion due to
mismatches in CSA and comparator design, the transition $\mu-$baseline median, which measures the
CSA output amplitude response to test pulse injection \Vtp, remains tightly distributed with a
mean value consistent with the expectation $\Vtp\cdot(\Cinj/C_f)$.

We write a set of values into the SRAM in each pixel to drive the 4-bit DAC to trim (reduce) the
threshold differences between pixels in the array.  The set of DAC values, $\{n_i\}$, are
calculated from the extracted parameters from threshold scans.  The threshold of pixel $i$ is
determined by $\Vth{_i}=\Vthg+\Vthp{_i}$.  All the 4-bit DACs share a globally adjustable step
size \Vs.  $\Vthp{_i}=\Vs\times n_i$.  Ideally, \Vth should be as close to the baseline median
while kept above the baseline noise width, to detect minimal signal amplitudes.  This requirement
points to a small \Vs.  However, at the same time, \Vthp should cover a maximal threshold
dispersion of the array in order to reduce the number of dysfunctional pixels due to insufficient
trimming.  Since $n_i$ has only 16 values, it points to a large \Vs, contracting the low
threshold requirement.  To find a balanced set of parameters, we minimize the quadratic sum of
signal thresholds, $\sum\limits_i(\Vth-\text{baseline median})_i^2$, by varying $\{n_i\}$, \Vs
and \Vthg.  We allow a small fraction of pixels with baselines that are far off to be excluded
and subsequently disabled.  We also disable defective and noisy pixels by setting $\sym{Mask}=0$.
Disabled pixels are marked with black points in the relevant 2D-figures.  A representative set of
parameters are $\Vs=\SI{9}{mV}$, $\Vthg=\SI{532}{mV}$, and $\SI{10}{\%}$ disabled pixels.

After trimming with the optimized setting, we varied \Vthg to perform the threshold scan again.
The results show a greatly reduced width in baseline median distribution
(Fig.~\ref{fig:baseline_mid}).  The signal threshold, however, has a somewhat high mean value and
wide distribution (Fig.~\ref{fig:Threshold}).  Ideally, if the trimming were able to equalize all
the baselines, which would require an infinitely small \Vs, the signal threshold distribution
would have a width equal to that of the distribution of transition $\mu-$baseline median.  A
finite (large) \Vs widens the signal threshold distribution and raises its mean value.

\begin{figure}[!htb]
  \centering
  \includegraphics[width=\linewidth]{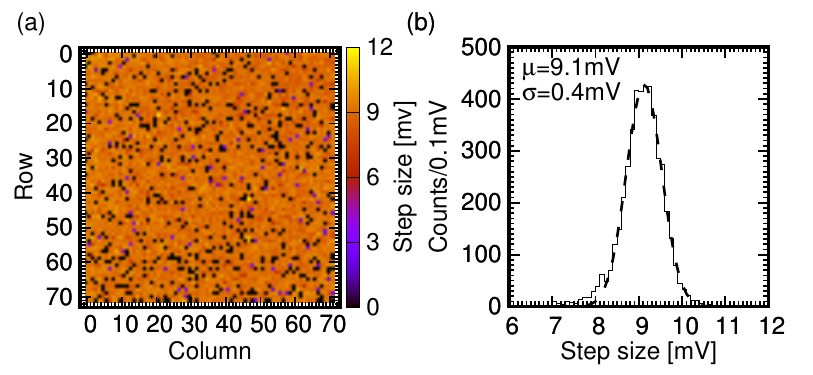}
  \caption{The step size distribution of in-pixel 4-bit DACs in 2D (a) and 1D histogram (b).
    Black points in the 2D distribution indicates the pixel is faulty and therefore its DAC is
    not testable.}
  \label{fig:DACStepDis}
\end{figure}

We also extracted the actual step size of the 4-bit DAC in each pixel.  The distribution is shown
in Fig.~\ref{fig:DACStepDis}.

\subsection{Imaging with pulsed LED illumination}

\begin{figure}[!htb]
  \centering
  \includegraphics[width=0.9\linewidth]{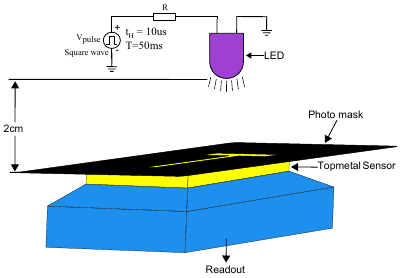}
  \caption{Light pulse injection setup.  A violet LED with a peak emission wavelength of
    \SI{390}{nm} is placed $\sim\SI{2}{cm}$ above the top surface of a \TMIIm sensor.  Light from
    the LED is filtered by a photo mask with a T-shaped transparent opening before arriving
    at the sensor.}
  \label{fig:setup}
\end{figure}

We placed a purple light LED $\sim\SI{2}{cm}$ above the top surface of a \TMIIm sensor.  The
sensor is covered by an opaque photo mask with a transparent T-shaped pattern.  The T-shaped
pattern is aligned with the center of the sensor (Fig.~\ref{fig:setup}).  The LED is driven by a
train of narrow pulses with \SI{10}{\micro s} width and \SI{50}{ms} interval.  The intensity is
set such that the illuminated pixels generate hits but their CSAs are not saturated.  A
$\sim\SI{1}{MHz}$ clock drives the CRUs and the MUX; therefore, the time it takes to read one
frame (all 72 columns for once) is $T_f\approx\SI{72}\times\SI{1}{\micro s}=\SI{72}{\micro s}$.
The width of the LED pulse is chosen to be narrow enough to be within one frame.  The interval
between pulses is large enough to allow all hits to be read out and all pixels to be reset.  The
sensor operates at the optimized threshold settings.

\begin{figure}[!htb]
  \centering
  \includegraphics[width=\linewidth]{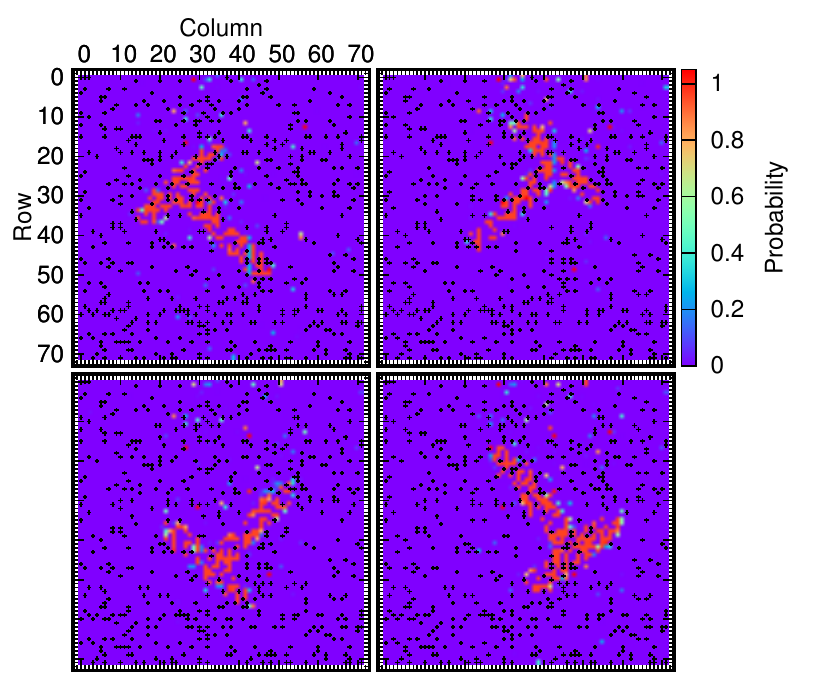}
  \caption{A set of images collected with the T-shaped photo mask placed at four different
    orientations.  Black points mark the disabled pixels.}
  \label{fig:T_four}
\end{figure}

We recorded many frames of hits induced by a large number of LED pulses.  The photo mask was also
rotated and displaced to cover different regions of the sensor.  Hit location and time are
reconstructed from data.  A set of images showing the T-shape at four different orientations is
in Fig.~\ref{fig:T_four}.

\begin{figure*}[!htb]
  \centering
  \includegraphics[width=\linewidth]{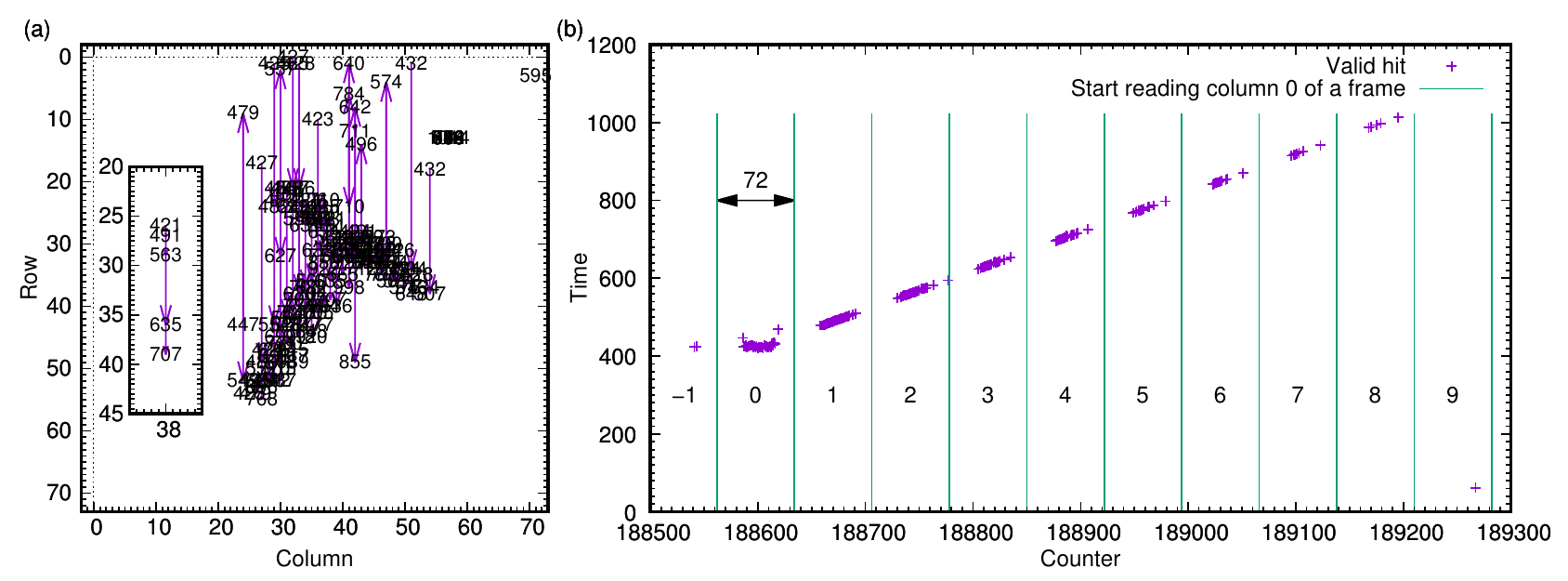}
  \caption{Time stamping of hits.  (a) Time stamps printed at the location of corresponding
    pixels resulting from one light pulse.  The T-shaped light hit pattern is clearly visible.
    Arrows connect hits in the same column, pointing from lower to higher time values.  (b) Time
    stamp (y-axis) read out as a function of clock cycle (Counter, x-axis).  Light pulse arrives
    in section (frame) $-1$.  Frame $0$ reads out the initial time stamps from highest priority
    pixels in each column.  Subsequent frames read out pixels hit by the same light pulse but
    with progressively lower priority.  Inset in (a) illustrates the readout of column No.~38 in
    this fashion.}
  \label{fig:TimeInfo}
\end{figure*}

When an LED light pulse arrives at the sensor, multiple active pixels that receive the light
generate a \sym{Hit} in each of them.  Due to the column readout logic, hit pixels that are in
the same column will have only one pixel that has the highest priority registered by the CRU.
Since CRUs from each column work concurrently while reading off a single globally shared time,
each CRU registers the hit time of the highest priority pixel in its column.  Since the light
pulse arrives at each pixel simultaneously, the initially registered time, which is from the
highest priority pixel, is the same for all the CRUs (Fig.~\ref{fig:TimeInfo} (a)).  The MUX
reads the registered time from each CRU in a round-robin fashion from one column to the next.
When a CRU is read, the pixel of the highest priority in its column is reset, and the CRU
subsequently registers the second-highest priority pixel.  Since only the CRU has access to the
global time, the hit time of the second-highest as well as all the lower priority pixels is
determined by the readout rather than the actual arrival of the signal.  Only the hit time of the
highest priority pixel is physically meaningful.  It is worth noting that starting from the
second-highest priority pixel, the time difference between the $i$th-priority pixel and the
$(i+1)$th-priority pixel in the same column equals the number of columns (72), which is the time
interval between consecutive reads for a given CRU (readout time for one full frame).
Fig.~\ref{fig:TimeInfo} (b) exemplifies this phenomenon.

\section{Summary and outlook}\label{sec:sno}

We successfully implemented a CMOS pixel sensor, \TMIIm, for direct charge collection and
imaging.  The detailed design, behavior and performance of a column-based priority logic readout
in the sensor are presented.  The electrical measurements and imaging applications demonstrated
the validity of such a readout scheme.  The digital readout of pixel hits features a fully
combinational logic in the pixel array and a sequential logic in the periphery.

In the current design, although the in-array combinational logic could drive the hit pixel's
address to the edge of the array with minimal latency, the sequential logic nature of the CRU and
the MUX limits the time it takes to discover the hit information to be beyond one clock cycle.
To further reduce the readout latency, analog and combinational logic could be designed at the
edge of the array to detect the activities in the Address Bus (AB) promptly.  A polling style MUX
could be replaced by a priority logic to read out the columns as well.  We will investigate these
options in future \TM sensor development in addition to improving the array uniformity.

\section*{Acknowledgments}

This work is supported, in part, by the Thousand Talents Program at CCNU and by the National
Natural Science Foundation of China under Grant No.~11375073.  We also acknowledge the support
through the Laboratory Directed Research and Development (LDRD) funding from Berkeley Lab,
provided by the Director, Office of Science, of the U.S.\ Department of Energy under Contract
No.~DE-AC02-05CH11231.  We would like to thank Christine Hu-Guo and Nu Xu for fruitful
discussions.



\bibliographystyle{elsarticle-num-names}
\bibliography{refs}

\end{document}